# Effective medium theory of conduction in stretched polymer electrolytes

Oliver Dürr, Wolfgang Dieterich and Philipp Maas
Fachbereich Physik, Universität Konstanz, D-78457 Konstanz, Germany

and

Abraham Nitzan
School of Chemistry, The Sackler Faculty of Science, Tel Aviv University,Tel Aviv 69978, Israel

09/08/01 18:42

## Abstract

Recent experimental observations of anisotropic conductivity in stretched polymer electrolytes films of the polyethylene oxide family are discussed. The main experimental observations, enhancement of the ionic diffusion and conductivity in the stretch direction and decrease in these transport coefficients in the normal direction are interpreted in terms of an effective two-phase model. This two-phase model is based on the idea that a highly conducting phase is associated with oriented molecular structures which are surrounded by poorly conducting boundary regions. This model is evaluated within the framework of differential effective medium theory (DEMT). Under stretching these regions change from spherical to prolate-spheroidal shapes. The computed dependence of the DC conductivity tensor and its AC counterpart on the stretch parameters is in good agreement with experimental results.

# 1. Introduction

Ionically conducting polymers such as polyethers containing alkali metal salts have been under investigation for more than two decades because of their unique combination of mechanical and electrical properties.[1,2] The continuing quest to improve the electrical transport properties of these materials is driven by their potential technological importance for electronic and energy storage devices. At the same time these materials have been subjects of intense theoretical interest focused on the mechanism underlying their ionic transport properties.[3]

Long time ago it has been proposed by Armand[4],[5] that the helical structure of polyether chains may provide a framework for ion transport in crystals of these materials. This suggestion as been subdued by the mounting evidence that long-range transport of ions is inhibited in crystalline phases of the host polymers and that ionic mobility in such systems is intimately connected to the host segmental motions (see, e.g. Ref. 1). Recent experimental and theoretical results have reopened the issue, also in conjunction with transport in the amorphous phase. First, using x-ray diffraction (XRD) spectroscopy, Bruce and coworkers have shown that in some stchiometries of crystalline polyethylene oxide (PEO) alkali cations are indeed enclosed within helical chains[6,7] or other directed structures[8,9] of the host, while being coordinated by oxygen atoms that belong to consecutive monomers in the host chain. Mechanisms for the temporal evolution of the cations' first coordination shells and the ensuing cation diffusion have been proposed on the basis of molecular dynamics simulations by Müller-Plathe et al.[10] These studies also indicate strong interactions between a chain and its ions and weak interchain interactions.[11] Secondly, in a series of recent articles Golodnitsky and coworkers[12-15] have shown that stretching films of PEO-Li complexes results in a profound effect on their ion transport properties. In particular the observed enhancement by more than an order of magnitude of the DC conductivity of such films along the stretch direction is consistent with a picture of ions moving along molecular chains (e.g. helices) that become preferably oriented along the stretch direction. This increased conductivity is accompanied by increasing stiffness, showing an apparent correlation between host stiffness and ion mobility that is contrary to what is commonly assumed.

Following is a more detailed summary of relevant experimental observations on these systems. The published results[12-14] are for LiI:PEO polymer electrolytes with $Li^+/O$ ratio 1:$n$, cast as films of width 300µm and length 15mm (before stretching) in the stretch direction. The sample thickness in the third perpendicular direction was about 8mm.

(a) Films with n=20 subjected to a load that exceeds 450-500N/cm$^2$ begin to 'flow' in the load direction. The film length in this direction increases by a factor of 3-6 while its thickness decreases by a factor of about 4. The DC conductivity in the stretch direction increases by a factor of 5-8 at $T$=40$^0$C and by factors 11-20 at $T$=60$^0$C. Both the stretching response to load and the resulting change in conductivity depend also on the ionic content measured by $n$, and for n=7 stretching at 60$^0$C was found[15] to cause conductivity increase by a factor of 40.

(b) $Li^+$ diffusion constants measured by the (Li-7) pulsed field gradient NMR technique show marked sensitivity to stretch.[15] The following results were obtained at $T$=60$^0$C for $n$=9 films that were stretched at the same temperature: $D(Li^+) \simeq 7.6 \cdot 10^{-8}$cm$^2$/s for the unstretched film and $D_{//}(Li^+) \cong 1.1 \times 10^{-7}$ cm$^2$/s; $D_{\perp}(Li^+) \cong 2.8 \times 10^{-8}$ cm$^2$/s for the stretched film (∥ and ⊥ refer to directions parallel and normal to the stretch, respectively). After the stretched film was annealed at 77$^0$C then cooled to $T$=60$^0$C, $D_{//}$ became $8 \cdot 10^{-8}$ cm$^2$/s.

(c) Scanning Electron Microscope (SEM) micrographs of the stretched samples show a fibrous-like structure in the stretch direction, indicating the formation of long-range structural anisotropy in the system. XRD data similarly indicate that oriented, presumably helical, microphases lie preferably along the stretch direction.

(d) Similar long-range anisotropy is seen from the angular dependence of the $^7$Li NMR spectrum. The unstretched sample shows no angular dependence of the linewidth while the stretched sample exhibits a pronounced $(3\cos^2\theta - 1)$-dependence ($\theta$ is the angle between the field and the stretch direction) that characterizes second-rank tensor interactions in systems with axial symmetry.

(e) The chemical shift in the $^7$Li NMR spectrum is considerably larger in stretched than in unstretched samples, indicating that short range effects, in this case a decrease in the nearest neighbor Li-O distance in the stretched sample, also exist.

(f) Infrared spectra change with stretching, again demonstrating short-range structural consequences of the configurational change.

(g) The stretched films appear to be stiffer than the unstretched samples. This is seen in the temperature dependence of the $^7$Li NMR linewidth. The onset of motional narrowing with increasing temperature, which is associated with enhanced segmental motion in the host polymer[16,17] is shifted to higher temperatures (by about 25K) in the stretched system relative to the unstretched one.[18] Similarly it has been recently observed[15] that the glass transition temperature of the $n=20$ system is shifted by 20K to higher temperature upon stretching. This increase in $T_g$ is considerably weaker in the concentrated rubber like systems with $n=7,9$.

(h) The spin lattice relaxation time ($T_1$) is longer in stretched than in the unstretched samples, reflecting differences in the local relaxation dynamics that again indicate short-range changes in the Li coordination. (It should be noted, however, that a shift of the minimum in the $T_1$ vs. $T^1$ plot to higher temperature, expected perhaps for the stretched samples if they are indeed more rigid, is not observed).

(i) Nyquist plots of the AC resistivity in the direction perpendicular to the stretch direction show marked differences between the stretched and unstretched samples. In particular the high frequency arcs of these plots, attributed in Ref. 14 to 'grain boundary resistance', including presumably the boundary between different molecular chains, are very different in the two cases. Room temperature conductivities of stretched LiI-$(PEO)_{20}$ and LiI-$(PEO)_7$ deduced from the low frequency intersections of these arcs with the real resistivity axis, are about half the corresponding values measured in the unstretched films.[15]

In this paper we present a theoretical model for the effect of stretching on the conduction properties of polymer electrolytes of the kind considered above. This model is based on the assumption that ionic conduction in such systems is governed by two mechanisms: One is associated with ion transport along directed molecular structures such as the helical chains in PEO. The other, strongly dependent on the host segmental motions, is controlled by ion hopping between such structures. Since the unstretched host is macroscopically isotropic, the average molecular shape in it is spherical by symmetry. Upon stretching, this average shape is distorted, becoming (again on the average) a prolate spheroid whose aspect ratio depends on the extent of the stretch. We show that this change in internal geometry leads to changes in conduction properties of the magnitude observed experimentally.

Following a presentation in Sect. 2 of a primitive model for the observed phenomenon we proceed in Sect. 3 to present the above model in more detail and suggest a variant of effective medium theory that seems to be best suitable for this model. In Section 4 we present numerical results for the behavior of the DC conductivity and discuss their applicability for the observations described above. Section 5 applies the suggested model to the analysis of the conductivity measured in AC experiments. While a simple extension of our model can account for these observations, applying effective medium theory to AC conduction of microscopically inhomogeneous media raises fundamental questions whose resolution still remains an open question. In Section 6 we conclude and outline the future research on this phenomenon.

## 2. A primitive model

Consider the simple model for stretching displayed in Fig. 1. This figure shows an *xy* projection of a sample in which conduction takes place by two mechanisms: a fast transport within the volume of each cell shown, and slow crossing across cell boundaries represented by the dividing lines in the figure. In what follows we use ∥ and ⊥ to denote the stretch ($x$) direction and the perpendicular ($y,z$) directions respectively, so that the original size of a unit cell is $d_x \times d_y \times d_z = d_\parallel \left( d_\perp \right)^2$. Upon stretching, under the restriction that the volume remains constant, these dimensions change to $d_\parallel \lambda$, $d_\perp / \sqrt{\lambda}$, where $\lambda$ is the parameter characterizing the stretch. The sample cross sectional area $A_\alpha$ perpendicular to the $\alpha$ direction scales like $A_\parallel \sim \lambda^{-1}$, $A_\perp \sim \sqrt{\lambda}$, and the sample length $D_\alpha$ in the $\alpha$ direction changes like $d_\alpha$. Under the assumption that the volumes within the cells conduct well and that the crossing of cell boundaries are rate determining, carrier transport may be thought of as a hopping process controlled by the slow crossing rates. In this picture stretching changes the displacement of an elementary hopping event in the corresponding direction, while the elementary hopping time is not affected. The conductivities therefore scale with $\lambda$ like

$$\sigma_\parallel \sim \lambda^2 \ ; \quad \sigma_\perp \sim \lambda^{-1} \tag{1}$$

while the conductances $G_\alpha = A_\alpha \sigma_\alpha / D_\alpha$ are unaffected. Note that this simple picture also predicts (trivially) that $\sigma_y$ and $\sigma_z$ are affected in the same way by the stretching in the $x$ direction. More important, it predicts that $\sigma_\parallel \sigma_\perp^2$ is not affected by this stretching. In Sect. 3 we shall come back to this primitive model, which already accounts for several main features, e.g. a nearly constant conductance $G_{//}$ observed in the experiment. A more detailed description, however, requires a refined modeling.

## 3. Effective medium model

As described in the Introduction, a variety of experimental methods have revealed structural ordering effects on different length scales in streched PEO samples. These include a preferred orientation of micron-sized domains as well as ordering on the molecular level, namely alignment and modification of the helical PEO-structure. Fast ion motion along helical chains, as suggested in Refs.[6,8,7], will then facilitate ion diffusion and enhance long-range conduction parallel to the stretch direction, but this process obviously cannot be rate-determining because long-range conduction implies transfer of ions between different chains as well as transport across inhomogeneities introduced by the more macroscopic domain structure.

In what follows we consider a simplified picture of this situation which combines those two aspects, fast transport inside aligned oriented regions of finite spatial extent and inefficient transport between those regions. To implement this picture we introduce an effective two-phase model which in the unstretched state consists of a dense distribution of highly conducting spheres which are embedded in a low conductivity "host" medium. Finally, stretching the polymer film will be represented in this coarse-grained picture by stretching the spheres, transforming them into prolate spheroids whose long axis lies in the stretch direction. For our subsequent treatment the actual identification of those regions with structural units in the PEO-sample is unimportant because the effective medium theory used below is not sensitive to the length scale associated with the inhomogeneous conductivity. Note that on this level of treatment it is irrelevant whether cations or anions are the more mobile charge carriers.

Clearly, the macroscopic properties of the system described will depend not only on the relative volume fractions of the differently conducting phases but also on their topological arrangement. To account for the latter characteristics we note that a similar approach has been successfully used by Cohen and coworkers[19,20] to describe conduction properties of sedimentary rocks.[21,22] The conductivity of such rocks results from interconnected water (i.e. conducting electrolyte solution)-filled pores in the rock, while the rock itself is a non-conducting solid. Because the pore space in such rocks remains interconnected down to very low values of the porosity $\phi$, there is no percolation threshold as a function of $\phi$. Our system is geometrically similar. The molecular chain entities (represented in our models by the spheres that transform to prolate spheroids upon stretching) correspond to the rock grains, and the highly entangled and winding space between them stands for the water-filled pores. The conduction properties interchange - in our case it is the space taken by the molecular entity that is highly conducting, while the intermolecular space is relatively insulating. Still, the methodology developed in Refs. 19,20 may be used. Effective medium theory and its differential version introduced by Cohen and coworkers are briefly reviewed next.

A calculation of the effective response of an inhomogeneous medium to an external field starts by dividing this response into the average property and the fluctuations from it. The electric field, for example, is computed as a sum of the incident field propagating in a homogeneous medium of dielectric constant $\varepsilon_0$ and the fields scattered by the local fluctuations $\varepsilon(\mathbf{r})$-$\varepsilon_0$. The latter are calculated by multiple scattering theory. In a mixture of $n$ phases $\varepsilon(\mathbf{r})$ is characterized by the intrinsic dielectric function of the $i$-th phase and by the position and shape of the corresponding spatial regions and the internal interfacial boundaries. In the single site approximation one considers a single particle $i$ characterized by a dielectric function $\varepsilon_i$ and subjected to a local field $\mathbf{E}_i$ that has been averaged over the local configuration of all other particles, and calculates the macroscopic polarization in terms of the particle's polarizability and the local field. For a two-component system characterized by spherical geometry this yields the effective dielectric constant in the form (see, e.g., 22,19)

$$\varepsilon_e = \varepsilon_0 \frac{\left(1 + 2\sum_{i=1}^{2} f_i \frac{\varepsilon_i - \varepsilon_0}{\varepsilon_i + 2\varepsilon_0}\right)}{\left(1 - \sum_{i=1}^{2} f_i \frac{\varepsilon_i - \varepsilon_0}{\varepsilon_i + 2\varepsilon_0}\right)} \tag{2}$$

where $f_i$ ($i$=1,2) is the volume fraction of phase $i$. Here the background dielectric response $\varepsilon_0$ is not yet defined. For a system in which a small concentration of spherical particles of phase 2 are distributed in the host 1 it is reasonable to choose $\varepsilon_0 = \varepsilon_1$. This yields the Maxwell-Garnett[23] generalization of the Clausius-Mossotti-Lorenz-Lorenz approximation,[24] and is sometimes known as the average t-matrix approximation (ATA). If the two phases are equivalent, a self-consistent approximation is obtained by taking $\varepsilon_0 = \varepsilon_e$, leading to an equation for $\varepsilon_e$ due to Bruggeman,[25] $\sum_i f_i (\varepsilon_i - \varepsilon_e)/(\varepsilon_i + 2\varepsilon_e)^{-1} = 0$. Here the two phases are treated symmetrically, however, the topological structure of having spherical inclusions is still maintained. When this result is obtained as an approximation employed in multiple scattering theory it is referred to as the "coherent potential approximation" (CPA). The same result is obtained from "effective medium theory" (EMA) that starts from the requirement that the average response $<\varepsilon(\mathbf{r})\mathbf{E}(\mathbf{r})>$ (where $\varepsilon(\mathbf{r})$ and $\mathbf{E}(\mathbf{r})$ are the local dielectric tensor and electric field, respectively, and where $<>$ denotes an ensemble average over the system disorder) is given by $\varepsilon_e<\mathbf{E}(\mathbf{r})>$, i.e.

$$\left\langle \left( \varepsilon(\mathbf{r}) - \varepsilon_e \right) \mathbf{E} \right\rangle = 0 \tag{3}$$

In what follows we will limit our considerations to the effective conductivity of the inhomogeneous medium. In this case the symmetric EMA reads[25]

$$\sum_{i=1}^{2} f_i \frac{(\sigma_i - \sigma_e)}{(\sigma_i + 2\sigma_e)^{-1}} = 0 \tag{4}$$

which can be derived from the analog of (3)

$$\left\langle \left( \sigma(\mathbf{r}) - \sigma_e \right) \mathbf{E} \right\rangle = 0 \tag{5}$$

Eq. (5) implies that the local fluctuation in the current density, associated with the deviation of the local conductivity from the effective conductivity, vanishes on the average. When applied to ellipsoidal particles with conductivity tensor $\sigma_i$, Eq. (5) leads to the following generalization of Eq. (4)[26]

$$\sum_{i=1}^{2} f_i \left( \sigma_i - \sigma_e \right) \left[ 1 + \sigma_e^{-1/2} \Lambda_i \sigma_e^{-1/2} \left( \sigma_i - \sigma_e \right) \right]^{-1} = 0 \tag{6}$$

where $\Lambda_i$ is a matrix associated with the phase $i$ that depends in general on $\sigma_e$. It is defined as follows: Let the matrix $\mathbf{B}$ define the ellipsoidal shape according to

$$\sum_{\alpha,\beta=1}^{3} B_{\alpha\beta} x_\alpha x_\beta = 1 \tag{7}$$

Define the matrix $\mathbf{R}$ to affect the transformation that diagonalizes the matrix $\boldsymbol{\sigma}_e^{1/2}\mathbf{B}\boldsymbol{\sigma}_e^{1/2}$ according to

$$\left(\mathbf{R}^{-1}\boldsymbol{\sigma}_e^{1/2}\mathbf{B}\boldsymbol{\sigma}_e^{1/2}\mathbf{R}\right)_{\alpha\beta} = \delta_{\alpha\beta} l_\alpha^{-2} \tag{8}$$

so that $l_\alpha$ are the corresponding eigenvalues. Finally, define the diagonal matrix $\mathbf{L}$ by $L_{\alpha\beta} = \delta_{\alpha\beta} L_\alpha$ where the "depolarization factors" $L_\alpha$ ($\alpha$=1,2,3) are given by:

$$L_\alpha = (1/2)P(0)\int_0^\infty \frac{d\mu}{\left(l_\alpha^2+\mu\right)P(\mu)}$$

$$P(\mu) = \left[\prod_{\alpha=1}^{3}\left(l_\alpha^2+\mu\right)\right]^{1/2} \tag{9}$$

Then

$$\boldsymbol{\Lambda} = \mathbf{R}\mathbf{L}\mathbf{R}^{-1}. \tag{10}$$

For spherical geometry we have $\mathbf{B} = \mathbf{R} = \mathbf{1}$, $l_\alpha = \sigma_e^{-1/2}$ and $L_\alpha = \Lambda_\alpha = 1/3$ ($\alpha$=1,2,3). Eq. (6) then yields Eq. (4).

As already mentioned, this 'conventional' EMA is not suitable for describing the model outlined above for stretched polymer conductors for two related reasons. First, as already noted, it treats the two phases symmetrically, while in our picture the low-conductivity medium always surrounds regions of the highly conducting phase. Secondly, when $\sigma_2$=0, standard effective medium theory predicts a 'percolation threshold' at a finite volume fraction of the conducting phase, while again our model implies that in this limit the DC conductivity vanishes for all system compositions.

To overcome this problem, we employ the *differential effective medium approximation* (DEMA) of Cohen and coworkers. In this approximation the composite material and its conductivity are built in infinitesimal stages. Using the index 1 for the surrounding phase of poor conductivity and the index 2 for the highly conducting phase, at the k+1-th stage a small volume, $d\upsilon$, of phase 2 is added to the effective medium that was obtained after the $k$-th stage. The conductivity $\sigma_{e,k+1}$ of the effective medium of the $k$+1-th stage is then computed using Eq. (6) with $\sigma_1$ replaced by $\sigma_{e,k}$, and the volume fractions $f_2$ and $f_1$ replaced by $d\upsilon/(\upsilon_1+\upsilon_{2,k})$ and $1-d\upsilon/(\upsilon_1+\upsilon_{2,k})\cong1$, respectively.[27]

Expressing the resulting effective conductivity $\sigma_{e,k}$ in terms of the volume fraction $f_{2,k}$ of phase 2 after the $k$-th step by

$$f_{2,k} = \frac{v_{2,k}}{v_1 + v_{2,k}} \tag{11}$$

leads for the following differential equation for $\sigma_e$

$$\frac{d\boldsymbol{\sigma}_e}{df_2} = \frac{1}{1-f_2}(\boldsymbol{\sigma}_2 - \boldsymbol{\sigma}_e)\left[1 + \boldsymbol{\sigma}_e^{-1/2}\boldsymbol{\Lambda}\boldsymbol{\sigma}_e^{-1/2}(\boldsymbol{\sigma}_2 - \boldsymbol{\sigma}_e)\right]^{-1} \tag{12}$$

This equation is integrated as an initial value problem, with the initial condition $\boldsymbol{\sigma}_e(f_2 = 0) = \boldsymbol{\sigma}_1$, up to the desired value of $f_2$.

## 4. Results and discussion

It is important to keep in mind the limitations of our model resulting from its coarse grained nature as well as from the heuristic way in which we introduced a two-phase system. This makes it difficult for a given material to assign volume fractions and local conductivities to those phases. For this reason we cannot hope to account quantitatively for any observation. Our aim is therefore to check whether the model introduced above can account for the qualitative observations in a consistent way.

Again $\parallel$ and $\perp$ denote the directions parallel and perpendicular to the stretch direction, respectively. In particular $\sigma_{e,\parallel}(\lambda)$, $\sigma_{e,\perp}(\lambda)$ and $\sigma_e = \sigma_{e,\perp}(\lambda = 0)$ $= \sigma_{e,\parallel}(\lambda = 0)$, denote the effective conductivities computed in the parallel and perpendicular directions, and the conductivity of the unstretched sample, respectively. As before, $\sigma_1$ and $\sigma_2$ denote the conductivities of the surrounding phase and of the ellipsoidal phase, respectively, and $\sigma_e$ is the computed effective conductivity. The stretch parameter $\lambda$ is the ratio between the long and the short axis of the ellipsoids,[28] with the long axis lying in the $\parallel$ direction. We display conductivities in relative units, taking $\sigma_2 = 1$.

Figures 2a-c show the effective conductivity computed for the isotropic case (Fig. 2a) and for the stretched sample with $\lambda$=4 (Fig. 2b for $\sigma_{e,\parallel}$; Fig. 2c for $\sigma_{e,\perp}$), as function of the volume fraction $f_2$ of the enclosed phase and for several values of $\sigma_1$. In all cases $\sigma_e$ approaches $\sigma_1$ for $f_2 \to 0$. For $f_2 > 0$ $\sigma_{e,\perp}$ varies with $f_2$ similarly as $\sigma_e$ in the unstretched sample, while the behavior of $\sigma_{e,\parallel}$ is quite different. Note in particular the change in the curvature, which implies that, for a given stretching $\lambda$, $\sigma_{e,\parallel}$ increases very strongly with $f_2$ for small $f_2$. Fig. 3 shows the ratios $\sigma_{e,\parallel}(\lambda)/\sigma_e$ (right panel) and $\sigma_{e,\perp}(\lambda)/\sigma_e$ (left panel) as functions of the stretch parameter $\lambda$ for the choice $\sigma_1 = 10^{-4}$ and for several values of $f_2$. In an intermediate range of $f_2$-values, $0.7 \leq f_2 \leq 0.9$, the initial slopes $(d\sigma/d\lambda)_{\lambda=1}$ are of the order predicted by the primitive model of Sect. 2, see Eq. (1). Indeed, some similarity between DEMA-theory and the model of Sect. 2 is to be expected for small $\sigma_1$ and large volume fractions $f_2$, as long as the poorly conducting regions dominate the overall conduction. Note that this last condition obviously requires parameters $\sigma_1$ and $f_2$ such that $\sigma_e(\lambda = 1) \ll 1$. This is satisfied with $\sigma_1 = 10^{-4}$, $f_2 = 0.9$, but is no longer fulfilled when $f_2 = 0.95$, as seen from the lowest curve in the right panel of Fig. 3, which deviates from the expected proportionality of $\sigma_{e,\parallel}$ to $\lambda^2$.

Clearly, the present model is more complex than the primitive model which does not include the volume fraction of the better conducting phase as a parameter, while in contrast the present model, when comparing results for different values of $f_2$, shows a correlation between a stronger dependence on $\lambda$ in the parallel direction and a weaker dependence on $\lambda$ in the perpendicular direction. Moreover, in the primitive model both $\sigma_{e,\perp}$ and $\sigma_{e,\parallel}$ scale with the inverse of the "hopping time" $\tau$ that is governed by the low conductivity $\sigma_1$. Accordingly, the primitive model cannot make predictions how the stretching effect on $\sigma_{e,\perp}$ and $\sigma_{e,\parallel}$ is influenced by the conductivity ratio between the two phases. The DEMA, by contrast, includes $\sigma_1/\sigma_2$ as a parameter. The dependence of the effective conductivities $\sigma_{e,\perp}(\lambda)$ (left panel) and $\sigma_{e,\parallel}(\lambda)$ (right panel) on this parameter is shown in Fig. 4. We see that for decreasing $\sigma_1$, the enhancement of $\sigma_{e,\parallel}$ with stretching becomes stronger than the corresponding reduction of $\sigma_{e,\perp}$.

We should emphasize at this point that the experimental work to date has not provided sufficient data about the dependence of the observed NMR and transport properties on the stretch parameter $\lambda$, and that such data is highly needed in order to assess the merit of the model presented here. Preliminary experimental data[29] does show a superlinear dependence of $\sigma_{e,\parallel}$ on the stretch parameter $\lambda$ after the polymer film commences its flow.

In agreement with experimental observations[15], our model show that stretching the polymer sample enhances the conductivity in the stretch direction while conduction in the perpendicular directions is inhibited. The magnitude of the effect, and the detailed way in which $\sigma_\parallel$ and $\sigma_\perp$ depend on $\lambda$ are sensitive both to the volume fraction $f_2$ and the conductivity ratio $\sigma_1/\sigma_2$. As detailed in Sect. 1, the experimental results of Refs. 14 and 15 point to another way in which stretching may affect the conduction properties of the polymer electrolyte - by increasing the stiffness of the polymer chain thereby inhibiting segmental motion. In our model this will reduce the conductivity of the "surrounding medium". To demonstrate this possible effect we show in Fig. 5 $\sigma_\parallel$ and $\sigma_\perp$ as functions of $\lambda$, both for the model used in Fig. 3 for volume fraction $f_2$=0.86 and for the same model modified so that $\sigma_1$ is assumed to scale like $\lambda^{-1}$ while $\sigma_2$ remains as before. We see that introducing such a dependence of $\sigma_1$ on $\lambda$ enhances the difference between $\sigma_\parallel$ and $\sigma_\perp$: both are reduced relative to their values in Fig. 4, but $\sigma_\perp$ is understandably affected more strongly than $\sigma_\parallel$.

Finally consider the frequency dependent conductivity. Golodnitsky and coworkers[14] have presented Nyquist plots for $\left(\sigma_\perp(\omega)\right)^{-1}$ for stretched and unstretched PEO/LiI films (O:Li = 20:1), that show a marked dependence on stretching. Similar studies in the parallel direction have not yet been done. In order to model the dispersive properties of the effective conductivity in our model we assume that the complex dielectric response of phases 1 and 2 may be written as

$$\varepsilon_i(\omega) = \varepsilon_i^\infty - \frac{4\pi i}{\omega}\sigma_i \tag{13}$$

and compute the components of the effective dielectric response tensor, $\varepsilon_{e,\parallel}(\omega)$ and $\varepsilon_{e,\perp}(\omega)$ from Eq. (12) with $\sigma$ replaced by $\varepsilon$ everywhere. The frequency dependent conductivity is then calculated from

$$\sigma_{e,\alpha}(\omega) = -\frac{\omega}{4\pi i}\left(\varepsilon_{e,\alpha}(\omega) - \varepsilon_{e,\alpha}(\infty)\right) ; \quad \alpha = \parallel, \perp \tag{14}$$

Figures 6 and 7 depict the dispersion properties of the resistivity computed by our model. We use dimensionless units defined by $\varepsilon_1^\infty = 1$ and $\sigma_1 = 1$ so that $\omega$ is measured in units of $\sigma_1/\varepsilon_1^\infty$. In these figures we have used $\varepsilon_2^\infty = 1$ and $\sigma_2 = 10^{-4}$ and volume fraction $f_2$=0.86. Fig. 6 shows the real and imaginary parts of the complex impedance, $z_{e,\alpha}(\omega) = \sigma_{e,\alpha}^{-1}(\omega)$ ; $\alpha = \|, \perp$ plotted vs. frequency for the unstretched polymer and for the stretched polymer with $\lambda=2$. Fig. 7 shows the Nyquist plots, Im($z$) against Re($z$) for the same systems. The slightly depressed semicircles associated with the unstretched polymer and with the stretched polymer in the perpendicular direction show the same trends observed in the experimental plots.[14] In particular, the low frequency (right) cutoff of these semicircles show again the effect of stretching on the DC resistivity, while the high frequency (left) cutoff is not sensitive to the stretching. We should keep in mind that the AC response of such systems may be influenced by factors such as boundary effects and inhomogeneous distribution of crystalline and amorphous phases that are not contained in our model, so final judgment concerning the interpretation of the available data should remain open until more experimental results are available.

## 5. Final Remarks

In this paper we have presented a simple model that qualitatively accounts for recent observations of the conduction properties of stretched polymer electrolyte films. The essential characteristics of our model are (a) the existence of two transport processes in ionically conducting polymers of the PEO type: a relatively fast process whose spatial extent is finite, and a slow rate determining process that connects between the regions covered by the fast process, and (b) the distortion of the latter regions when the polymer sample is macroscopically stretched. Other details of our model, e.g. describing the stretching-induced anisotropy as a transition from spherical to spheroidal shapes of the highly conduction regions, are helpful for the mathematical description but should not be considered critical components of the theory. In particular, we note that an alternative model that takes the highly conducting regions to be spheroids (or cylinders) whose *orientational distribution* changes from isotropic to non-isotropic upon stretching is probably more appropriate for cases, where, e.g., oriented microphases already exist without stretch and provide the main cause of a heterogeneous conductivity. It would be interesting to consider this alternative (and technically more involved) model in future work.

Other assumptions made in our model remain to be tested by future experiments. Controlled measurements of the film conduction properties as functions of the stretch parameter $\lambda$ in the parallel and perpendicular directions will be critical in this respect. In this context it is important to mention again the difference that may exist between the macroscopic and microscopic distortions. Macroscopic distortion always implies a change of shape. We have assume that such shape change takes place also on the microscopic molecular scale, but one can envision a later stage of the stretch process in which the macroscopic shape changes due to redistribution of already elongated structural units in space, without further changes in their shape. Again, such possibilities should be considered in future work if warranted by further experimental data.

**Acknowledgements**. This research was supported by the Lion Foundation and (AN) by the Israel Science Foundation. We thank D. Golodnitsky, E. Peled and S. Greenbaum for many helpful discussions and for communicating to us their experimental results prior to publication. Philipp Maass thanks the Deutsche Forschungsgemeinschaft for financial support by a Heisenberg fellowship.

(18)    It should be kept in mind that [7]Li-based data on this issue are not conclusive, because at least part of the $Li^+$ dynamics, i.e. its motion along the stretch direction is enhanced by the stretching. The present interpretation is however consistent with the observed increase in $T_g$ upon stretching. [1]H NMR data would provide additional support.

(26)    Eq. (6) is in fact a special case of a more general expression applicable for many phases (each associated with an ellipsoidal geometry), where random orientations of the ellipsoids are possible. In the present paper these options are not considered.

(27)    This process has been formulated, following Ref. (20), as adding phase 2 into the effective system, so that the total volume $v_{tot,k} = v_1 + v_{2,k}$ changes while the volume of phase 1 remains constant. Alternatively one may construct the final structure by a different route, e.g., a process in which the total volume $v_{tot} = v_{1,k} + v_{2,k}$ remains constant. The final result, Eq. (12), does not depend on the choice of route.

(28)    We assume that the macroscopic stretch is fully reflected in the microscopic stretch, i.e., the macroscopic distortion measured by $\lambda$ is taken equal to the microscopic aspect ratio of the spheroids. This is not necessarily true. In fact the experimental results of Ref. (14) indicate that $\lambda_{macro}$ and $\lambda_{micro}$ are related in a way that is somewhat dependent on the sample's temperature.

(29)    Golodnitsky, D., et al. private communication.

FIGURE CAPTIONS

<u>Figure 1.</u> A primitive model for the effect of stretching on transport. The system is divided into cells that are distorted by the stretching as shown. The process of crossing the cell boundary lines is assume to be rate determining.

<u>Figure 2.</u> The effective conductivities of the unstretched film (a) and of a stretched film with stretch parameter $\lambda$=4. (b-conductivity in the stretch direction, c-conductivity in the normal direction) as functions of the volume fraction $f_2$. Full line: $\sigma_1$=0.1, Dashed line: $\sigma_1$=0.01, Dotted line: $\sigma_1$=10$^{-4}$.

<u>Figure 3.</u> The ratio between the conductivities of the stretched film in the stretch direction (right panel) and in the perpendicular direction (left panel) and the conductivity of the unstretched film as functions of the stretch parameter $\lambda$ for several values of the volume fraction $f_2$ and $\sigma_1$=10$^{-4}$. The initial slopes at $\lambda$=1 are, in the order of increasing $f_2$ $-(d\sigma_{e,\perp}/d\lambda)_{\lambda=1}$ = 0.59, 0.88, 0.89, 0.85, 0.68 and $(d\sigma_{e,\parallel}/d\lambda)_{\lambda=1}$ = 1.14, 2.00, 2.03, 1.98, 1.49.

<u>Figure 4.</u> Same as Fig. 3, now for different values of $\sigma_1$ ($\sigma_2$ is taken 1). Note that the full lines ($\sigma_1$=10$^{-4}$) here are identical to the dashed lines ($f_2$=0.86) in Fig. 3.

<u>Figure 5.</u> The parallel and perpendicular conductivities of the stretched sample as functions of the stretch parameter $\lambda$ for the original model (full lines), and for a model in which $\sigma_1 \sim \lambda^{-1}$ (dashed lines). $\sigma_1(\lambda$=1) is taken 10$^{-4}$. The lines with positive slopes correspond to $\sigma_{e,\parallel}$ and the others - to $\sigma_{e,\perp}$.

<u>Figure 6.</u> The real (top) and imaginary (bottom) components of the complex impedence z($\omega$) of the isotropic ($\lambda$=1; full line) and the stretched samples ($\lambda$=2; dotted and dashed lines correspond to $\sigma_{e,\parallel}$ and $\sigma_{e,\perp}$, respectively) plotted against the frequency $\omega$.. $f_2$=0.86 and $\sigma_1$=10$^{-4}$ were used in this calculation.

<u>Figure 7.</u> The imaginary part of the complex resistivity z plotted against its real part for the unstretched sample (full line) and for the stretched sample ($\lambda$=2) in the parallel (dotted line) and the perpendicular directions (dashed line). Other parameters are as in Fig. 6.

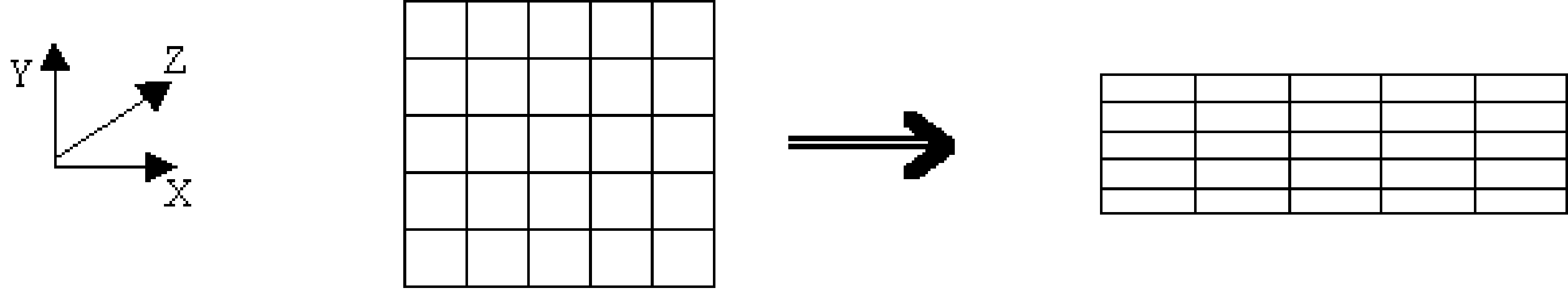

FIG. 1

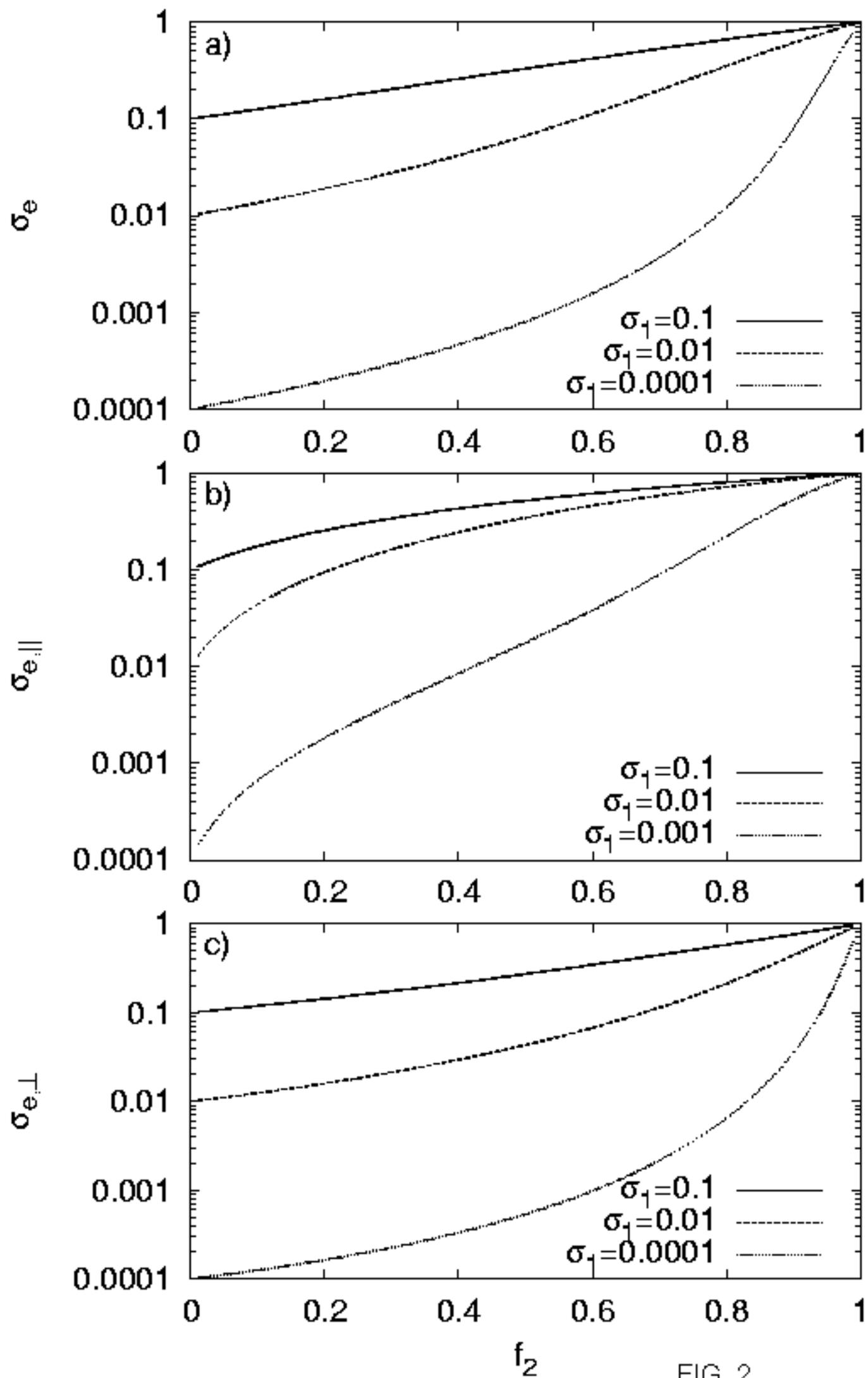

FIG. 2

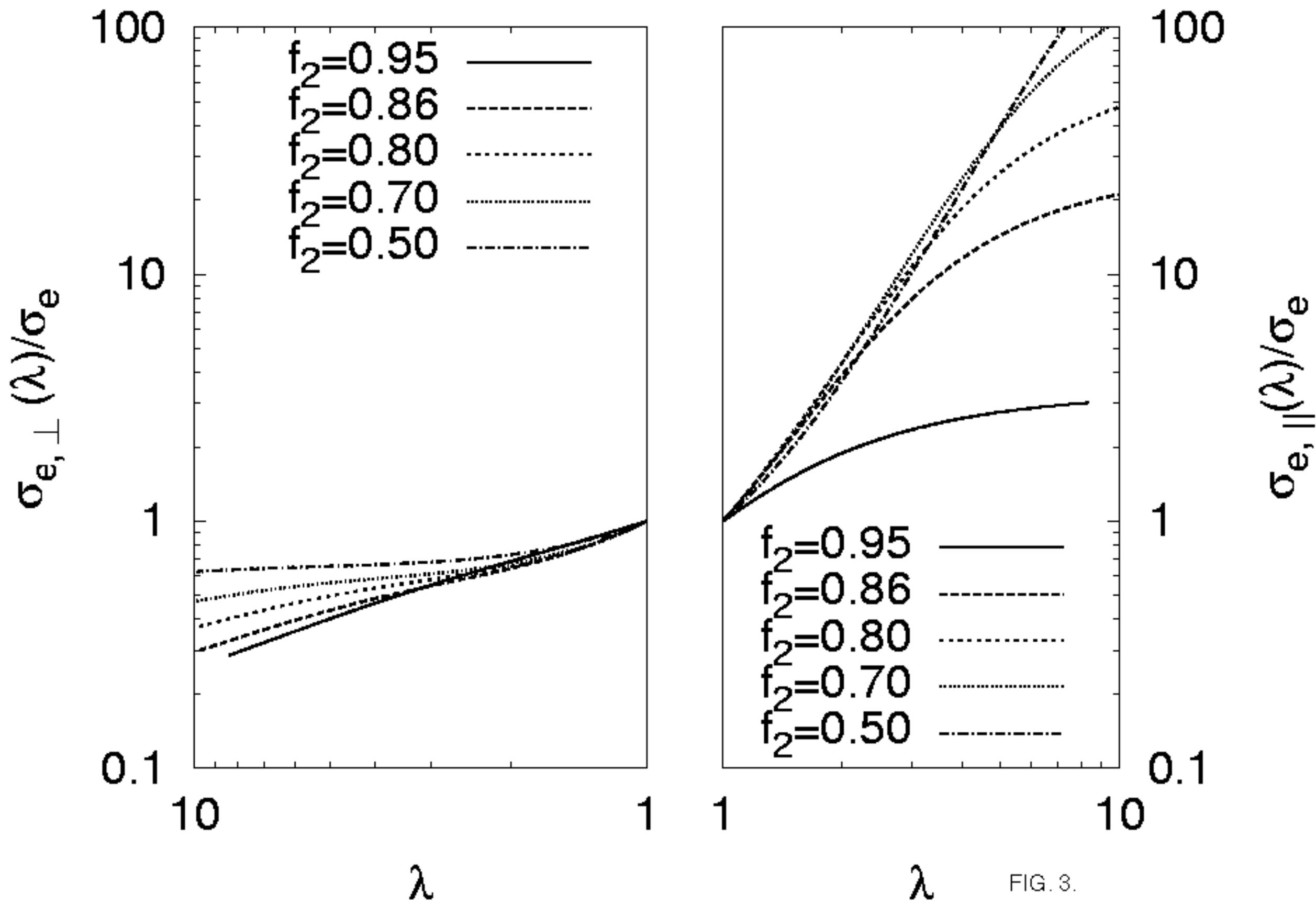

FIG. 3.

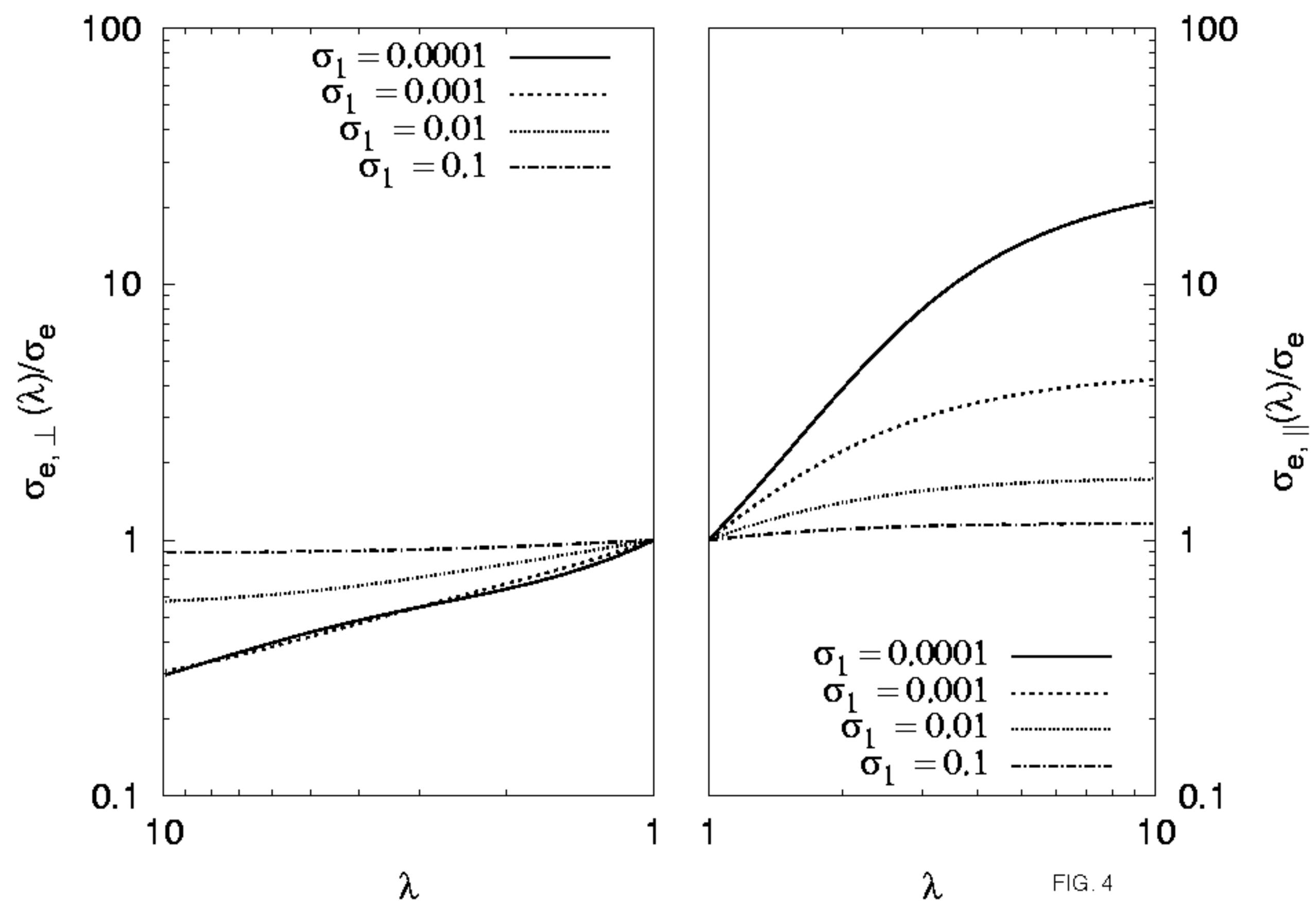

FIG. 4

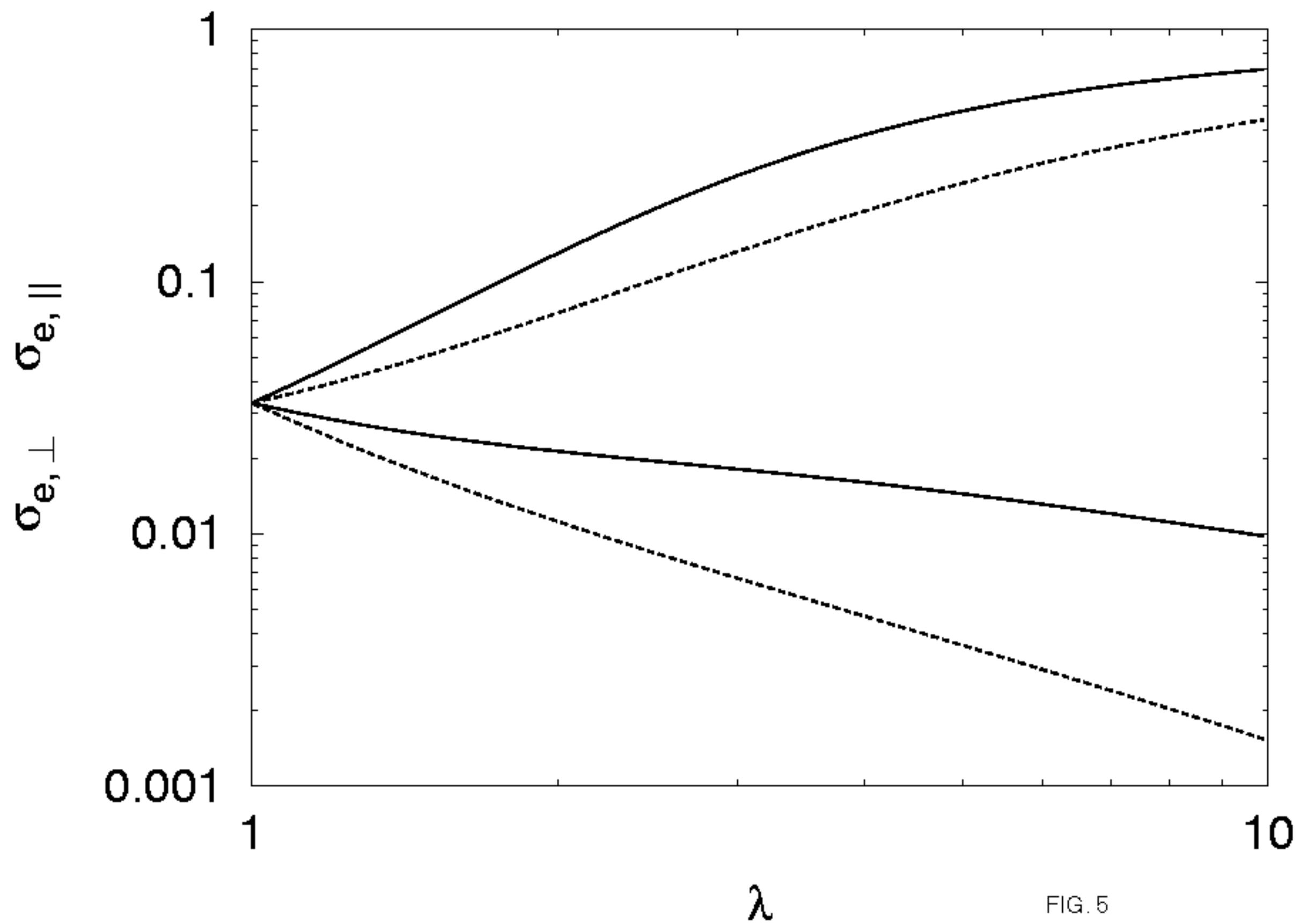

FIG. 5

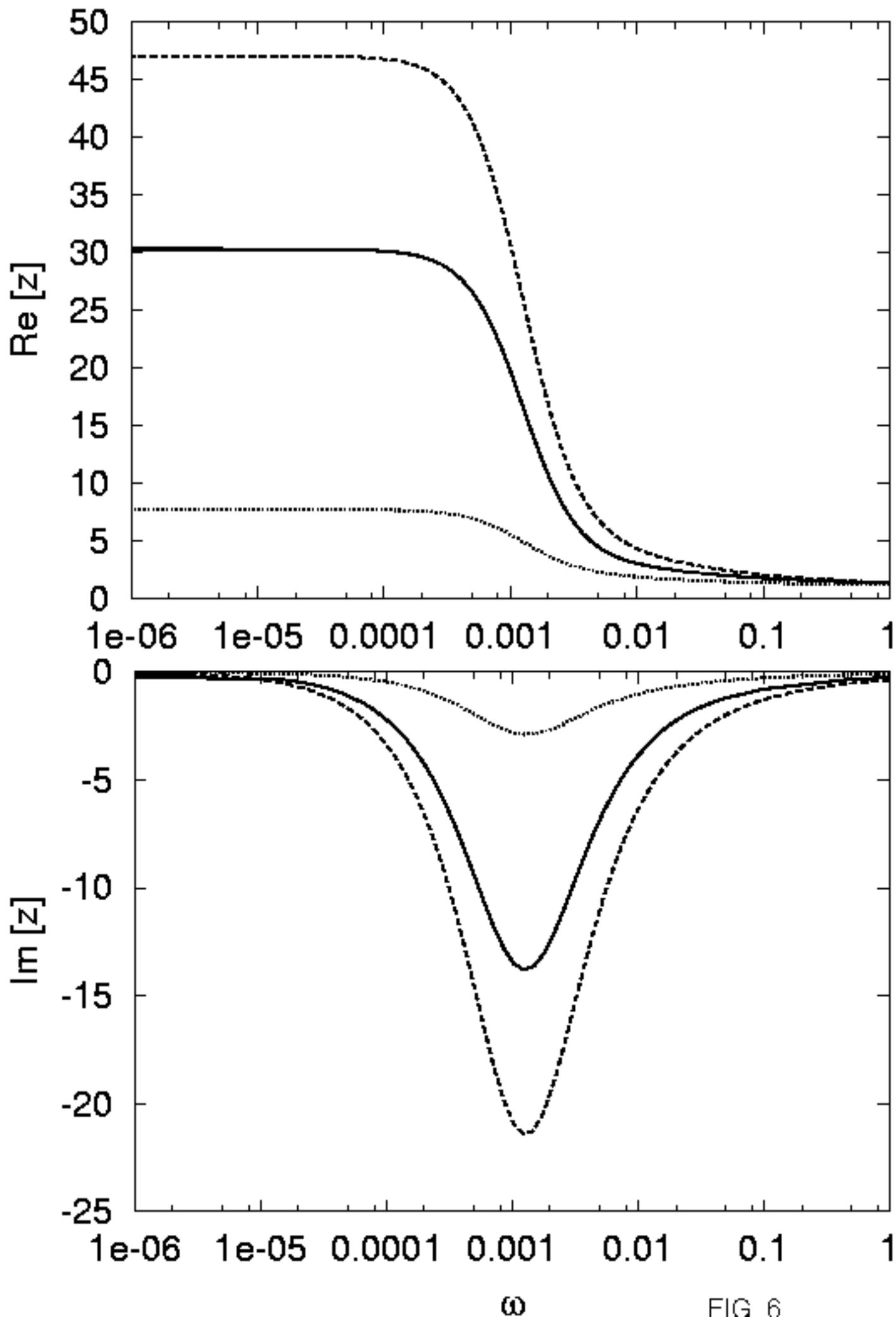

FIG. 6

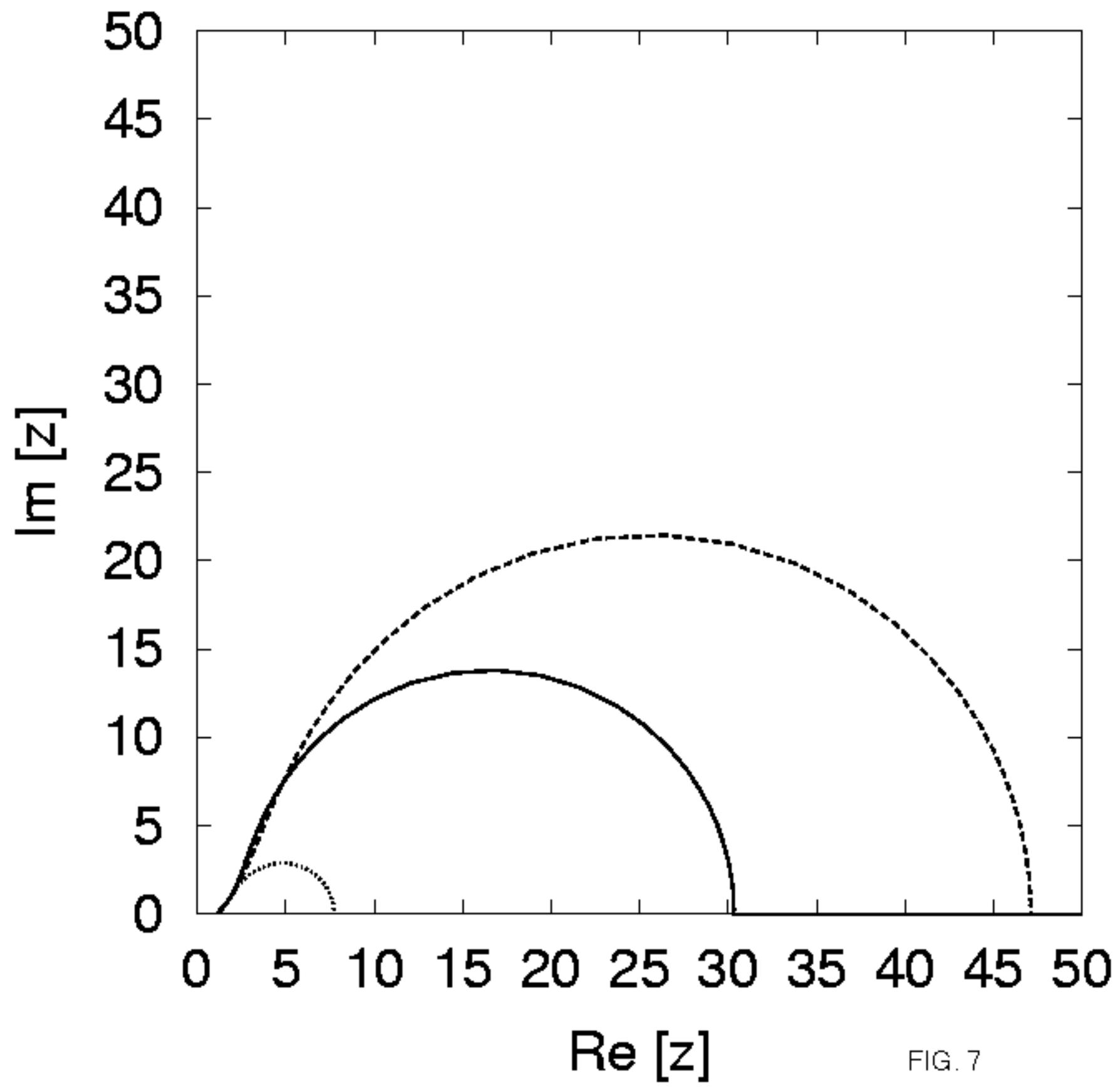

FIG. 7